\documentclass[lettersize,journal]{IEEEtran}
\usepackage{amsmath,amsfonts}
\usepackage{algorithmic}
\usepackage{algorithm}
\usepackage{array}
\usepackage[caption=false,font=normalsize,labelfont=sf,textfont=sf]{subfig}
\usepackage{textcomp}
\usepackage{stfloats}
\usepackage{url}
\usepackage{verbatim}
\usepackage{graphicx}
\usepackage{cite}
\usepackage[columnwise]{lineno} 
\usepackage{xcolor}  

\begin{document}

\title{Contrastive Learning-Based Agent Modeling \\ for Deep Reinforcement Learning}
\author{Wenhao Ma, Yu-Chen Chang, Jie Yang, Yu-Kai Wang,~\IEEEmembership{Member,~IEEE}, and Chin-Teng Lin,~\IEEEmembership{Fellow,~IEEE}
\thanks{The authors are with the Australian AI Centre, School of Computer Science, Faculty of Engineering and Information Technology, University of Technology Sydney, Ultimo, NSW 2007, Australia (e-mail: wenhao.ma@student.uts.edu.au; yu-cheng.chang@uts.edu.au; jie,yang-1@uts.edu.au; yukai.wang@uts.
edu.au; chin-teng.lin@uts.edu.au).
\\© 2025 IEEE.  Personal use of this material is permitted.  Permission from IEEE must be obtained for all other uses, in any current or future media, including reprinting/republishing this material for advertising or promotional purposes, creating new collective works, for resale or redistribution to servers or lists, or reuse of any copyrighted component of this work in other works.}
}



\maketitle

\begin{abstract}
Multi-agent systems often require agents to collaborate with or compete against other agents with diverse goals, behaviors, or strategies. Agent modeling is essential when designing adaptive policies for intelligent machine agents in multi-agent systems, as this is the means by which the controlled agent (ego agent) understands other agents' (modeled agents) behavior and extracts their meaningful policy representations. These representations can be used to enhance the ego agent's adaptive policy which is trained by reinforcement learning. However, existing agent modeling approaches typically assume the availability of local observations from modeled agents during training or a long observation trajectory for policy adaption. To remove these constrictive assumptions and improve agent modeling performance, we devised a \textbf{C}ontrastive \textbf{L}earning-based \textbf{A}gent \textbf{M}odeling (\textbf{CLAM}) method that relies only on the local observations from the ego agent during training and execution. With these observations, CLAM is capable of generating consistent high-quality policy representations in real time right from the beginning of each episode. We evaluated the efficacy of our approach in both cooperative and competitive multi-agent environments. The experiment results demonstrate that our approach improves reinforcement learning performance by at least 28\% on cooperative and competitive tasks, which exceeds the state-of-the-art. 
\end{abstract}


\section{Introduction}
\IEEEPARstart{T}{he} rapid evolution of artificial intelligence has led to the widespread deployment of machine agents in real-world settings, where they collaborate with other machines and humans as multi-agent systems to achieve diverse tasks \cite{o2022human}. In multi-agent systems, intelligent agents are expected to work tacitly with agents having diverse goals, behaviors, or strategies. Maintaining effective and flexible responses when facing different types of agents in a multi-agent environment becomes a crucial question \cite{yuan2023learning}. This challenge leads the agent modeling as a key research topic within the field of multi-agent systems \cite{shoham2007if,albrecht2018autonomous}.
Agent modeling, which entails developing computational models to infer and comprehend the behaviors and policies of different agents, plays a significant role in creating intelligent machine agents with adaptive policies for multi-agent systems. \cite{yuan2023survey}
Specifically, it enables the controlled agent (ego agent) to acquire a more profound comprehension and prediction of the policies of other agents (modeled agents) present in the system. This capacity can be utilized to augment the decision-making process and empower the ego agent to adapt its behavior in accordance with the policies of the modeled agents. For this reason, agent modeling, as a kind of policy representation technique, has become an indispensable and vital component of multi-agent systems.

Existing approaches \cite{he2016opponent,grover2018learning,liam} involve using observation signals from both the ego agent and the modeled agents to generate distinctive policy representations through deep neural network models. These representations, commonly referred to as ``context information", are then combined with the ego agent's real-time observation signals to train adaptive policies for cooperating or competing with other agents via reinforcement learning\cite{sutton2018reinforcement}.
\cite{grover2018learning} leveraged the encoder-decoder architecture to learn modeled agents' policies by reconstructing the observation of the ego agent. However, it relies on long historical observation trajectories from the ego agent to generate informative policy representation. LIAM \cite{liam}, following the same architecture, feeds the ego agent's local observations into the encoder during the training phase and reconstructs the local observation of the modeled agent. This process builds the predictive capability of the ego agent regarding the policies of modeled agents, which is why the LIAM model solely relies on the ego agent's local observation during execution. However, the assumption that observations from the perspective of modeled agents are available during the training phase may not always hold in many real-world scenarios. Especially in human-autonomy collaboration, it is difficult for ego agents to obtain observational information from a human's perspective for agent modeling tasks.

\begin{figure}
\includegraphics[width=0.42\textwidth]{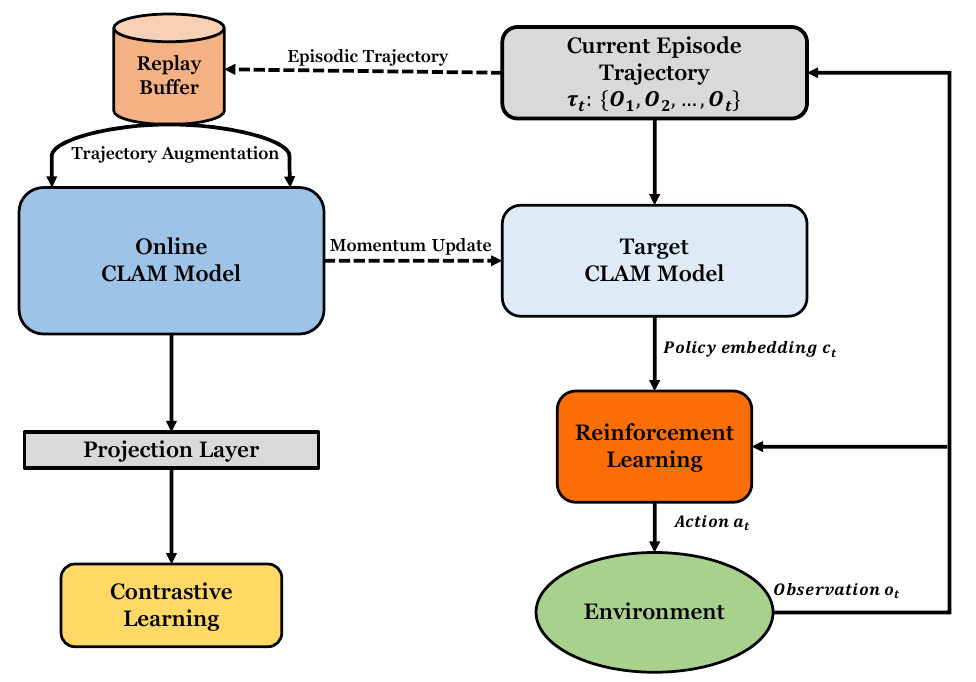}
\caption{Diagram of our proposed CLAM model and adaptive policy training architecture. The left part is for CLAM training. The right part is for adaptive policy training.} \label{fig1}
\end{figure}

To address these challenges, we propose a self-supervised representation learning method to achieve agent modeling. Employing both attention mechanism \cite{vaswani2017attention} and contrastive learning \cite{simclr}, the method encodes the ego agent's local observation trajectories using an attention-based policy encoder and identifies the most representative trajectory features by an asymmetric sample augmentation strategy in contrastive learning. Through this process, our method produces differentiated policy representation embeddings of modeled agents based on the interaction between the ego agent and modeled agents, which creates a more straightforward and practical solution for real-world scenarios. The key contributions of our method are as follows:

\begin{itemize}
    \item We propose a Contrastive Learning-based Agent Modeling (CLAM) \footnote{Code available at \url{https://github.com/WenhaoMa-UTS/CLAM-RL}}
 model that consists of a Transformer encoder module and an attention pooling module. The employment of the attention mechanism in two modules enables CLAM to dynamically allocate weights to different parts of the observation sequence data based on their importance. This process significantly enhances the model's capability to capture the most representative feature elements during the feature representation and aggregation stages.

    \item The model is trained using self-supervised contrastive learning with accompanying asymmetric sample augmentation strategy that creates positive sample pairs. The results of comparative experiments demonstrate that our method produces better policy representations than those produced through symmetric sample augmentation techniques. Moreover, this novel method enables the model to generate policy representations of modeled agents in real time and only requires local observations of ego agents.

    \item The unified training architecture concurrently trains the agent modeling model and reinforcement learning model resulting in a more concise, efficient, and easily deployable framework, as shown in Fig. \ref{fig1}.
\end{itemize}

\section{Related Work}

\subsection{Agent Modeling}
Agent modeling (or opponent modeling) plays a crucial role in the design of adaptive policies for intelligent machine agents within multi-agent systems, as it enables the ego agent to acquire a deeper comprehension and prediction of the policies of other agents (modeled agents) present in the system. This capacity can be utilized to augment the decision-making process and empower the ego agent to adapt its behavior in accordance with the policies of the modeled agents.  \cite{he2016opponent} proposed a neural network-based model to jointly learn the adaptive policy and agent modeling model. The Theory of mind Network (TomNet) \cite{rabinowitz2018machine}, used two networks called character net and mental net to infer the goals of modeled agent efficiently. \cite{grover2018learning} proposed an encoder-decoder architecture, which embeds the input observation trajectory into feature vectors in a point-wise manner and uses the average pooling method to generate the representation of the trajectory. \cite{liam} also based on an encoder-decoder architecture. However, they use a recurrent encoder which can better leverage historical observation information to aid the ego agent. These agent modeling approaches either assume the availability of local observations from other agents (modeled agents) during training or a long observation trajectory for policy adaption. The key distinction between our approach and these studies lies in that we only rely on observations from the ego agent and adapt the ego agent's policy within a few steps.

\subsection{Representation Learning in Reinforcement Learning}
In the field of reinforcement learning, representation learning typically serves as an auxiliary task to transform high-dimensional, complicated observation signals into low-dimensional, informative, and compact feature representations, thereby enhancing the efficiency and effectiveness of policy training in complex environments. Prior research has explored diverse methodologies to achieve meaningful representations in reinforcement learning scenarios. \cite{van2016stable} trains a variational autoencoder to learn task-related state features from high-dimensional sensor data. This reconstruction-based representation learning method maps high-dimensional states to a low-dimensional state representation space, thereby improving the sample efficiency and training stability of reinforcement learning. 
CURL \cite{laskin2020curl} is the first to utilize contrastive learning to train a visual representation encoder capable of extracting compact state features from raw pixel inputs. In \cite{zhu2022masked}, the author introduces M-CURL, which leverages both contrastive learning and masked pre-training to improve the existing CURL method. It also incorporates a Transformer module to capture correlations among consecutive frames in video data. Temporal abstraction, another significant aspect, has been tackled through hierarchical reinforcement learning, enabling agents to discover hierarchies in states and actions for improved learning efficiency \cite{precup2000temporal}. Meta-learning techniques have also been explored, aiming to facilitate policy adaption across tasks or domains \cite{wang2021improving}. These advancements collectively reflect the ongoing efforts to unlock the potential of representation learning in reinforcement learning, aiming to accelerate convergence, boost exploration, and facilitate more effective policy learning.

\subsection{Contrastive Learning}
In recent years, contrastive learning has emerged as a prominent paradigm in the realm of machine learning and representation learning. The underlying principle of contrastive learning revolves around learning representations by maximizing the similarity between positive pairs and minimizing the similarity between negative pairs \cite{cpc}. This concept has gained widespread interest and has been extensively explored across a spectrum of domains, encompassing computer vision and natural language processing \cite{jaiswal2020survey}. Seminal works like SimCLR \cite{simclr} have paved the way by introducing the concept of large-batch training coupled with negative samples to augment the quality of learned representations. Building upon this foundation, MoCo \cite{moco} incorporated a momentum moving-averaged encoder to facilitate better feature extraction. Additionally, approaches like SwAV \cite{caron2020unsupervised} have ventured into novel augmentation strategies to enhance contrastive learning. 

\begin{figure*}
\begin{center}
\includegraphics[width=0.85\textwidth]{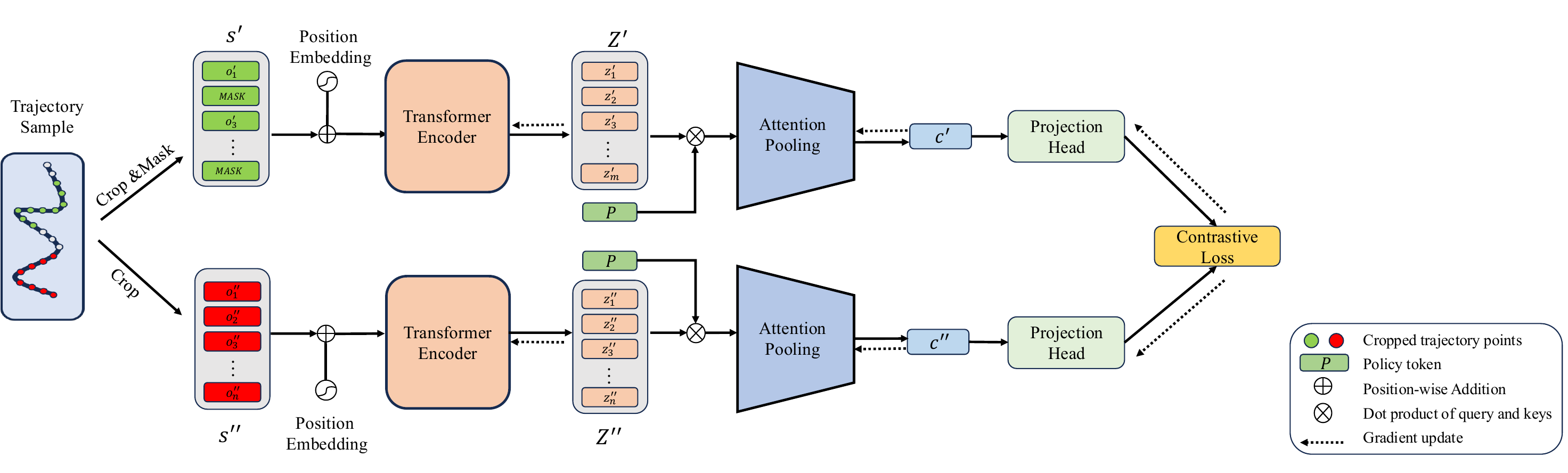}
\caption{Diagram of CLAM model training process.} \label{fig2}
\end{center}
\end{figure*}

\section{Method}
\subsection{Problem Statement}
In this section, we formally introduce the problem to be addressed.
Overall, our goal is to create an intelligent machine agent (ego agent) that uses its local observations to identify the intentions, behavioral preferences, or policies of other agents (modeled agents) within the same environment. These characteristics are encoded into representation vectors through an agent modeling model. Subsequently, leveraging these representation vectors, the ego agent generates corresponding policies to interact with the modeled agents through reinforcement learning training.

This entire problem can be modeled as a Partially Observable Stochastic Game (POSG) \cite{hansen2004dynamic} involving $N$ agents, defined as the tuple $(I, S, A, O, P, R)$. Here, $I = \{1,...,n\}$ represents the set of agents. $S$ represents the set of states, $A = A_1 \times A_2 \times … \times A_N$ is the joint action space and $O = O_1 \times O_2 \times … \times O_N$ is joint observation space. $P$ represents the state transition and observation probabilities, where $P(s_{t+1}, o_{t+1}| s_t, a_t)$ denotes the probability of taking joint action $a_t$ at state $s_t$ to the next state $s_{t+1}$ and joint observation $o_{t+1}$ and $R$ represents the reward function, where $R: S \times A \rightarrow \Re$.

To ensure clarity and precision, the local observations of the modeled agent and the ego agent at time t are denoted as $o^m$  and $o^e$, respectively. It is assumed that a set of predefined fixed policies exists $\Pi = \{\pi_i^m | i = 1, 2, …, n\}$, which can either be hand-coded or created via reinforcement learning. The policies within this policy set are randomly selected and made available for execution by the modeled agent. A fixed policy $\pi_i^m(a^m_t | o^m_t)$ represents a function that maps the modeled agent's current observation $o^m_t$ to a distribution over its action $a^m_t$. Our objective is to train the ego agent's adaptive policy $\pi_\alpha$, which is parameterized by $\alpha$, to achieve the maximum expected return \footnote{``return" refers to the total accumulated reward the agent receives over a sequence of steps.} through interactions with the policies in the fixed policy set $\Pi$:

\begin{equation}
    \mathbb{E}_{\pi_i^m \sim u(\Pi)} [\mathbb{E}_{\pi_ {\alpha}, \pi_i^m } [\sum_{t=0}^{T-1} \gamma^tr_{t} (o^e_t, a_t)]] 
    \label{eq_1}
\end{equation} 

where the inner expectation accounts for the discounted cumulative reward acquired by the ego agent throughout one episode length $T$ and the outer expectation accounts for the ego agent's average episodic return over various policies $\pi_i^m$ of the modeled agent. The policy $\pi_i^m$ is drawn from a uniform distribution $u(\Pi)$. $r_t(\cdot)$ is the ego agent's reward function and $\gamma \in (0, 1)$ is the discount factor to balance the trade-off between immediate return and long-term return.

\subsection{Contrastive Learning-based Agent Modeling}
To address our research problem, a contrastive learning-based agent modeling model is proposed to generate real-time policy representation of modeled agents. As its input, this model takes an observation trajectory of the ego agent from the beginning of an episode $0$ up to the current time $t$, i.e., $\tau_t = [o^e_1, o^e_2, ..., o^e_t]$. It then generates and outputs a policy representation vector $c_t$. This vector is intended to encapsulate the policy-specific information of the modeled agent, including its objectives, its behavioral characteristics, and policy information. This real-time policy representation vector $c_t$ will be used in conjunction with the real-time observations $o^e_t$ of the ego agent to condition the adaptive policies as $\pi_\alpha(a_t|o^e_t, c_t)$.

The proposed CLAM model consists of two components: (1) a Transformer encoder to transform the original observation trajectory $\tau_t$ into corresponding feature representation sequence $Z_t:= \{z_i\}_{i=0:t}$, where $z_i$ is the representation vector at time-step $i$; (2) an attention pooling module to aggregate the temporal feature sequence $Z_t$ into a policy representation vector $c_t$. The following sections will provide a detailed description of the two components.

\textbf{Transformer Encoder:}  The observation trajectory of the ego agent can be considered a form of sequential data, which naturally prompted us to consider using a Transformer encoder $f_\theta(\cdot)$, parameterized by $\theta$, to capture patterns and relationships within the ego agent's observation trajectories. We leverage the standard Transformer encoder \cite{vaswani2017attention} which consists of multiple multi-head self-attention blocks. 
Firstly, we encode the ego agent's observation trajectory with position embeddings to retain temporal information. Given the ego agent's observation trajectory $\tau \in \mathbb{R}^{t \times d}$, where $d$ and $t$ represent observation vector dimension and trajectory length respectively, we encode the observation trajectory $\tau$ with position embeddings method $\textit{pos}(\cdot)$ to retain temporal information, i.e.,
\begin{equation}
    \tau\prime = \tau + \textit{pos}(\tau).
    \label{eq_2}
\end{equation}

Then, the multi-head self-attention method is used to extract policy features from different perspectives simultaneously, which has been found to significantly enhance the model's representation capacity \cite{devlin2018bert,radford2018improving,dosovitskiy2020image}. 
Given $n$ attention heads, $\tau\prime$ is projected into the query $\mathbf{Q}_n$, key $\mathbf{K}_n$ and value $\mathbf{V}_n$ as:
\begin{equation}
    \{\mathbf{Q}_n, \mathbf{K}_n, \mathbf{V}_n\} = \{\tau\prime \mathbf{W}^Q_n, \tau\prime \mathbf{W}^K_n, \tau\prime \mathbf{W}^V_n\},
    \label{eq_3}
\end{equation}
where $\mathbf{W}^Q_n \in \mathbb{R}^{d \times d_k}$, $\mathbf{W}^K_n \in \mathbb{R}^{d \times d_k}$, $\mathbf{W}^V_n \in \mathbb{R}^{d \times d_v}$, $d_k=d_v=d/n$ are learnable weight matrices.
The computation of multi-head attention can be expressed as follows:

For the $n$th head $h_n$:
\begin{equation}
    \textit{h}_n = \textit{Softmax}\left(\frac{\mathbf{Q}_n\mathbf{K}_n^T}{\sqrt{d_k}}\right)\mathbf{V}_n 
    \label{eq_4}
\end{equation}
Finally, the outputs of each head are concatenated and linearly transformed to obtain the observation trajectory's corresponding feature representation sequence $Z_t$ as:
\begin{equation}
    Z_t = \textit{concat}(h_1, h_2, \ldots, h_n)\mathbf{W}^{O}
    \label{eq_5}
\end{equation}
where $\mathbf{W}^{O}$ is a trainable weight matrix used for linear transformation.

\textbf{Attention Pooling Module:}  After the Transformer encoder generates the feature embedding sequence $Z_t$ for each episode step within the observation trajectory, the next step is to aggregate the feature embedding sequence into a single policy representation vector for the current interaction time-step. Conventional methods usually employ average pooling \cite{rabinowitz2018machine,grover2018learning} for this feature aggregation step. However, since each vector in the trajectory might have a different contribution to the final policy representation vector, we believe that using an attention mechanism might better capture complex temporal relationships and patterns within the trajectory. Thus, our model appends an attention pooling module $pool_\phi(\cdot)$ parameterized by $\phi$ after the Transformer encoder. We create a special trainable policy token $\textit{P}$ $\in$ $\mathbb{R}^{1 \times d}$ as the query in the attention mechanism to aggregate the feature embedding sequence $\textit{$Z_t$}$, attention pooling function $pool_\phi(\cdot)$ with policy token $\textit{P}$ is defined as:
\begin{equation}
    pool_\phi(\textit{$Z_t$}) = rFF(\textit{MultiHead}(\textit{P}, \textit{$Z_t$}, \textit{$Z_t$}))
    \label{eq_6}
\end{equation}
where $rFF$ is row-wise feedforward layers and $\textit{MultiHead}$ means the multi-head attention mechanism. This policy token $\textit{P}$ helps to aggregate the entire input feature vector sequence $\textit{$Z_t$}$ into a single policy representation vector $c_t$. Our experiments also verify that attention-based pooling helps the agent to get higher returns compared to average pooling. A detailed analysis can be found in the ablation study.

\textbf{Contrastive Learning Training:} We optimize the parameters of the transformer encoder $\theta$ and attention-based pooling module $\phi$ through self-supervised contrastive learning. The overall training process is depicted in Fig. \ref{fig2}.
Firstly, We select a batch of size $N$ observation trajectories $\{\tau_{i}\}_{i=1:N}$ from the replay buffer. Then, we randomly crop two trajectory windows from each sampled trajectory and form two sample sets $S=\{s_{i}\}_{i=1:N}$ and $S^{\prime}=\{s^{\prime}_{i}\}_{i=1:N}$. The two cropped samples $s_i$ and $s^{\prime}_i$ extracted from the same original trajectory $\tau_i$ can be considered positive sample pairs, while trajectory samples cropped from different original trajectory samples serve as negative sample pairs. Subsequently, we conduct secondary sample augmentation on one sample set $S$. This involves randomly masking a certain ratio of observation steps in the trajectory sample, resulting in a new sample set $S^{\prime \prime}$, while the other sample set $S^{\prime}$ remains unaltered. The masked observation steps are replaced with 0. The sample set $S^{\prime \prime}$ subjected to mask-based augmentation is termed ``strong augmentation," while the unaltered one $S^{\prime}$ is referred to as ``weak augmentation". This asymmetric sample augmentation approach has been experimentally shown to enhance the performance of contrastive learning models \cite{wang2022importance}. We believe the underlying rationale for our asymmetric sample augmentation approach is to compel the model to extract more representative policy features from sparser and less informative observation trajectories. This approach enhances the model's ability to capture subtle patterns, making it more adept at handling real-time policy representation tasks.

We utilize the InfoNCE loss proposed in \cite{cpc} as the loss function for contrastive learning:

\begin{equation}
\begin{aligned}
    \mathcal{L}_{\text{C}}(\theta, \phi) &= -\frac{1}{N} \sum_{i=1}^{N} \log\frac{\exp(\mathbf{c^{\prime}}_i \cdot \mathbf{c}_i^{\prime \prime} / \delta)}{\sum_{k=1}^{N}  \exp(\mathbf{c^{\prime}}_i \cdot \mathbf{c}_k^{\prime \prime} / \delta)},
    \label{eq_7}
\end{aligned}
\end{equation}
where $N$ is the batch size, $\delta$ is a temperature parameter, $c_i^{\prime}$ and $c_i^{\prime \prime}$ is a positive pair of policy representations and $c_k^{\prime \prime}$ is any other samples generated by the agent modeling model. 
Note that a projection head is also used, which is the same as in \cite{simclr}.

\begin{algorithm}[h]
\caption{CLAM Model and Adaptive Policy Training}
\label{alg:clam}
\begin{algorithmic}
\STATE \textbf{Initialize}: PPO actor network $\pi_\alpha$ and critic network $V_\beta$, PPO target actor network $\pi_\alpha'$ and target critic $V_\beta'$; CLAM encoder $f_\theta$ and pooling function $pool_\phi$, CLAM target encoder $f_\theta'$ and target pooling function $pool_\phi'$; Clam replay buffer $\mathcal{B}$; Modeled agent policy set $\Pi$; PPO update frequence $freq_{ppo}$; CLAM update frequence $freq_{CLAM}$; Number of training episode $N$;

\FOR{episode $=1$ to $N$}
    \STATE Reset the environment, create the episodic trajectory buffer $\tau_{episode}$, and sample modeled agents' policy $\pi_i^m \sim \Pi$;
        \FOR{t = 1 to  max-episode-length}
        \STATE Compute policy embedding: $c_t =pool_\phi'(f_\theta'(\tau_{episode}))$;
        \STATE Select ego agent action $a_t = \pi_\alpha(o_t^e, c_t)$;
        \STATE Select the modeled agent action: $a^m_t  = \pi_i^m(o^m_t)$;
        \STATE Execute joint actions $a = (a_t, a^m_t)$, obtain reward $r_t$ and next step observation $(o^e_{t+1}, o^m_{t+1})$;
        \STATE Store ego agent observation $o^e_{t+1}$ in $\tau_{episode}$;
        \ENDFOR
    \STATE Store episodic observation trajectory $\tau_{episode}$ in $\mathcal{B}$;
    \IF{\text{$episode\mod freq_{ppo} = 0$}}
        \STATE Update PPO actor network $\pi_\alpha$ and critic network $V_\beta$ based on Eq.~\eqref{eq_8} and Eq.~\eqref{eq_9}, respectively;
        \STATE Update target model parameter: 
        \STATE $\alpha' \gets \tau \alpha + (1 - \tau) \alpha'$, $\beta' \gets \tau \beta + (1 - \tau) \beta'$;
    \ENDIF
    \IF{\text{$episode\mod freq_{CLAM} = 0$}
            \STATE \AND {($|\mathcal{B}| \geq buffer\_capacity$)}}
        \STATE Sample batch trajectory $\{\tau_1, \cdot\cdot\cdot,\tau_n\}$ from buffer $\mathcal{B}$;
        \STATE Update CLAM encoder $f_\theta$ and pooling function $pool_\phi$ based on Eq. \eqref{eq_7};
        \STATE Update target model parameter: 
        \STATE $\theta' \gets \tau \theta + (1 - \tau) \theta'$, $\phi' \gets \tau \phi + (1 - \tau) \phi'$;
    \ENDIF
\ENDFOR
\end{algorithmic}
\end{algorithm}

\subsection{Reinforcement Learning Training}
Reinforcement learning is employed to train adaptive policies. The real-time output policy representation vector $c_t$ from the agent modeling mode is combined with the ego agent's observation state $o_t^e$, to condition the optimal policy, i.e., $\pi_{\alpha}(a_t|o_t^e, c_t)$. As a result, the trained agent policy can generate optimal action outputs specialized for the representation information $c_t$.
In this paper, we select Proximal Policy Optimization (PPO) \cite{schulman2017proximal} as the backbone algorithm for reinforcement learning training. However, it's important to note that other reinforcement learning algorithms can also be equally applicable. To further demonstrate CLAM's generalizability, we conduct an additional experiment based on Deep Q-Network (DQN) \cite{mnih2013playing}, a value-based reinforcement learning algorithm, the results of which are provided in the supplementary materials. The objective function of policy $\pi_{\alpha}(\cdot)$ is defined as follows: 
\begin{equation}
\begin{aligned}
\mathcal{J}^\text{PPO}(\alpha) = \mathbb{E} \bigg[ &\textit{min} \bigg( \frac{\pi_\alpha( a_t | o_t^e, c_t)}{\pi_{\alpha_\text{old}}(a_t | o_t^e, c_t)} \cdot A_t, \\
&\textit{clip}\left( \frac{\pi_\alpha(a_t | o_t^e, c_t)}{\pi_{\alpha_\text{old}}(a_t | o_t^e, c_t)}, 1 - \epsilon, 1 + \epsilon \right) \cdot A_t \bigg)\bigg] \\
&+ c \cdot \mathbb{E}\left[H(\pi_\alpha(a_t | o_t^e, c_t))\right],
\label{eq_8}
\end{aligned}
\end{equation}

where $A_t$ is the advantage function, representing the advantage of taking action $a_t$ given observation $o_t^e$ and real-time policy representation vector $c_t$, $H(\cdot)$ is the entropy function, $c$ is the coefficient of entropy regularization and $\textit{clip}(\cdot)$ is a clipping function used to restrict the magnitude of policy updates. The PPO's value function $V_\beta(\cdot)$ is optimized by minimizing the following objective function: 
\begin{equation}
\mathcal{L}^\text{PPO}(\beta) = \mathbb{E} \left[ (V_\beta(o_t^e, c_t) - R_t)^2 \right]
\label{eq_9}
\end{equation}
where $\beta$ is the parameter of the value function, $V_\beta(o_t^e, c_t)$ estimates the value at each time-step $t$ based on the ego agent observation $o_t^e$ and policy representation embedding $c_t$, $R_t$ is the expected return. The overall CLAM model and adaptive policy training algorithm is summarized in Algorithm \ref{alg:clam}.



\section{Experiments}
\subsection{Environments}
We selected three representative multi-agent environments to evaluate our method: a cooperative environment, level-based foraging \cite{christianos2020shared,papoudakis2020benchmarking}, a competitive environment, predator-prey \cite{mordatch2018emergence}, and a mixed cooperative and competitive environment, StarCraft Multi-Agent Challenge (SMAC) \cite{samvelyan2019starcraft}. We created a fixed policy set of ten policies for each environment. At the beginning of each episode, a policy is randomly selected from the policy set with equal probability and assigned to the modeled agent.

\begin{figure}[t]
\centering
\subfloat{\includegraphics[width=0.148\textwidth]{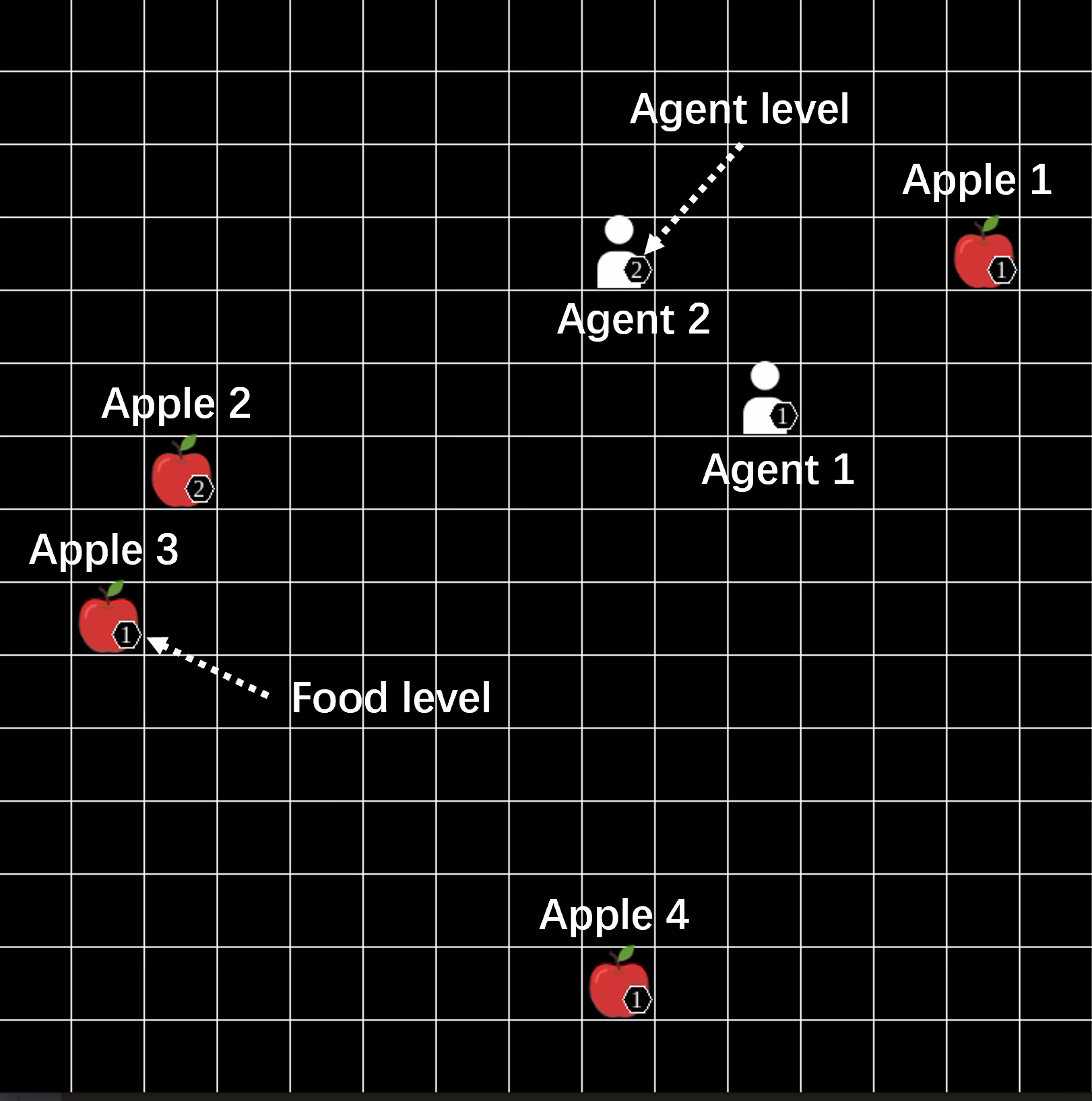}\label{fig3a}}
\hspace{0.1cm} 
\subfloat{\includegraphics[width=0.148\textwidth]{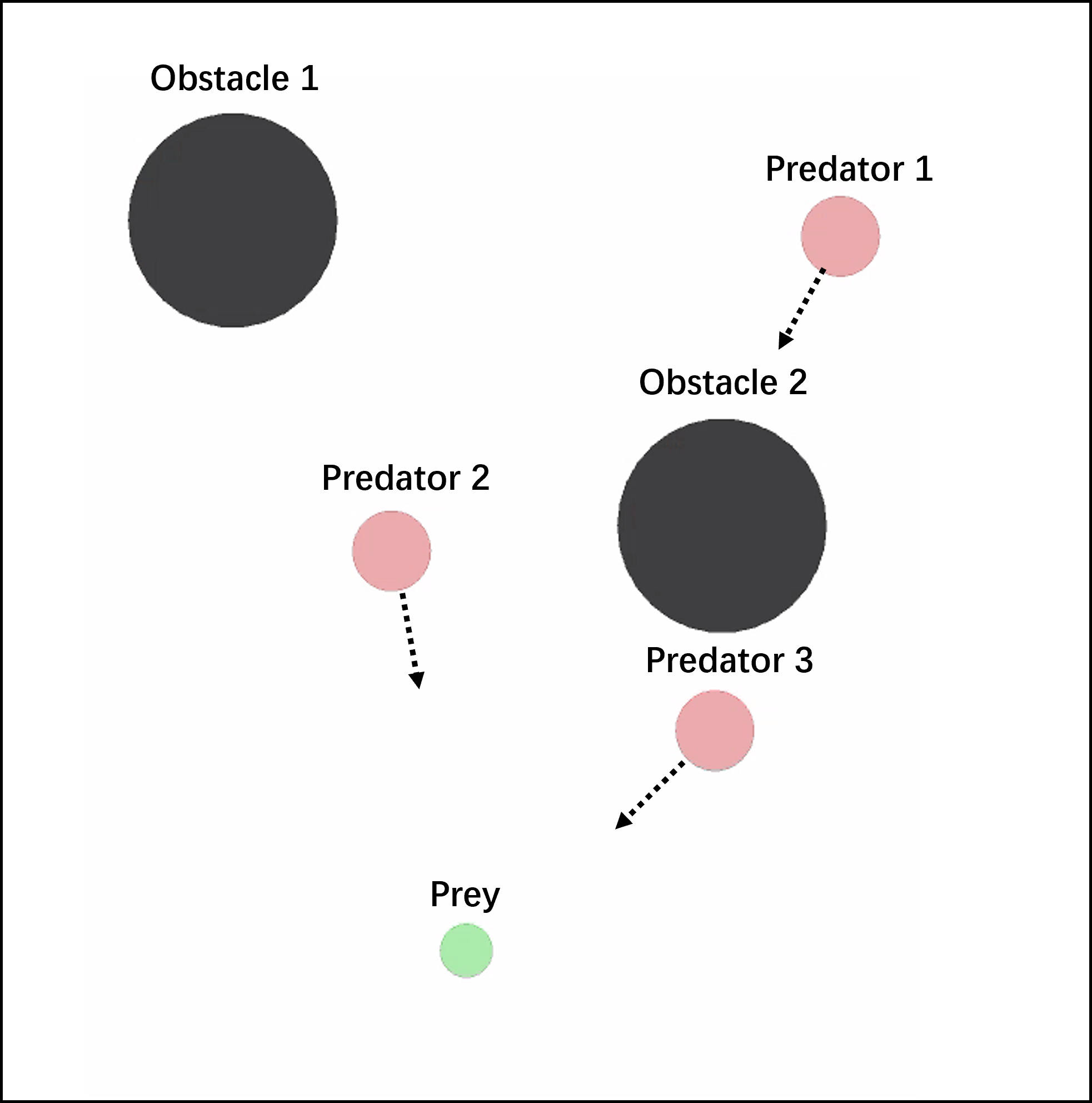}\label{fig3b}}
\hspace{0.1cm} 
\subfloat{\includegraphics[width=0.165\textwidth]{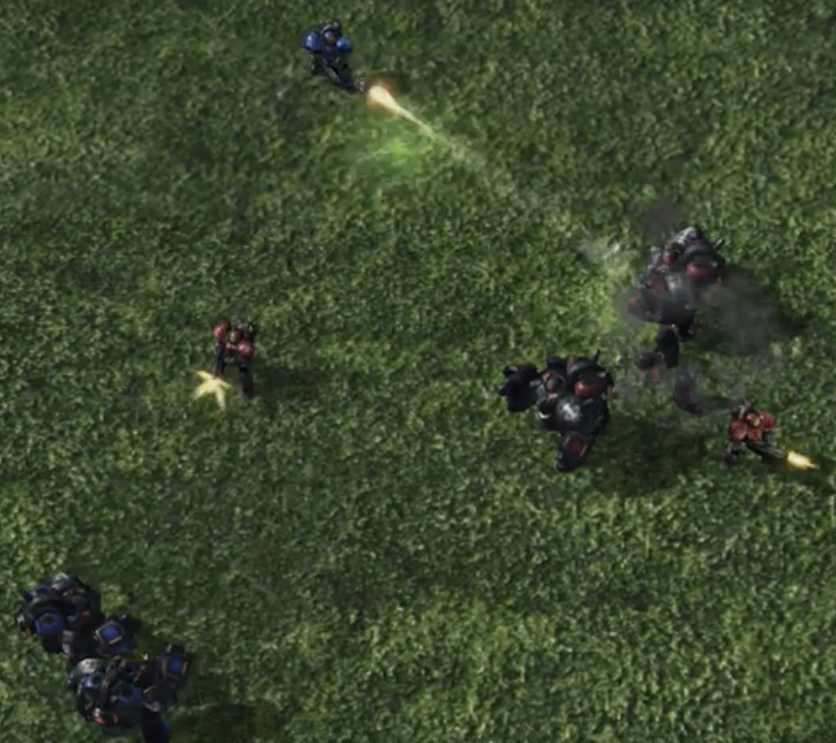}\label{fig3c}}
\caption{The three multi-agent environments for evaluation. (a) Level-based foraging.  (b) Predator prey. (c) SMAC}
\label{fig3}
\end{figure}

\paragraph{Level-based Foraging} 
This is a grid environment consisting of two agents and a number of apples, as shown in Fig. \ref{fig3a}. Each agent and apple has its level value, and the agent has six action options: up, down, left, right, stay still, and pick the apple. Agents can successfully pick up an apple when one or both agents are located in one of the four grid cells adjacent to the targeted apple and the sum of the agents' levels is not less than the apple's level. When an apple is picked up, agents receive rewards based on their contribution to completing the task. The total reward value for each episode is normalized to 1.

An episode ends when all apples are picked up or after 50 time steps. In this environment, one of the agents serves as the modeled agent while the other agent is the ego agent and is trained via reinforcement learning. Since our aim is to assess the model's performance in a cooperative environment, we compute the total reward obtained by both agents as the team returns in each episode. A higher value indicates more efficient cooperation among the agent team.

\paragraph{Predator-prey} 
This competitive multi-agent environment consists of three predators, one prey, and two circular obstacles, as illustrated in Fig. \ref{fig3b}. Each agent has five action options: up, down, left, right, and stay still. The three predators are assigned fixed policy combinations, while the prey is the ego agent under training. The reward setting in the environment is such that when a predator collides with the prey, the prey receives -10 and the predator receives +10. In the event of a predator-predator collision, each predator receives -3 and the prey receives +3. This reward scheme encourages the prey to learn adversarial policy by evading the predator team's pursuit while inducing more chaos and collisions among them. 

This environment is highly suitable for evaluating the effectiveness and performance of agent modeling methods. By employing agent modeling methods to acquire informative policy representations, the ego agent can learn targeted adversarial policy based on different opponent policies to obtain greater rewards. Each episode lasts for 50 time steps.
\paragraph{StarCraft Multi-Agent Challenge (SMAC)}
This widely recognized benchmark scenario is used to evaluate decentralized control tasks, where allied agents must learn to coordinate effectively to defeat enemy agents controlled by built-in AI, as depicted in Fig. \ref{fig3c}. Each agent has a limited-radius observation range and a discrete action space, including movement in four cardinal directions, stopping, attacking a specific enemy unit, and using special abilities if applicable. The agents receive team-based rewards that encourage cooperative behaviors.
In this scenario, one ally agent follows a fixed adversarial strategy, pick from a diverse adversarial policy set, which is generated using the evolutionary generation method proposed in ROMANCE \cite{yuan2023robust}, while the remaining allied agents must learn to cooperate to compete against the enemy team. This turns the environment into a mixed cooperative-competitive scenario. Each allied agent needs to infer the policy used by the adversarial teammate based on its local observation through agent modeling methods and adapt its cooperative policy accordingly to achieve victory. We evaluate model performance based on the ally team win rate, where a higher value reflects better coordination. We select the 3m and 2s3z maps to evaluate our algorithm.

\begin{figure}[!t]
\centering 
\subfloat{
\label{Fig4.sub.1}
\includegraphics[width=0.23\textwidth]{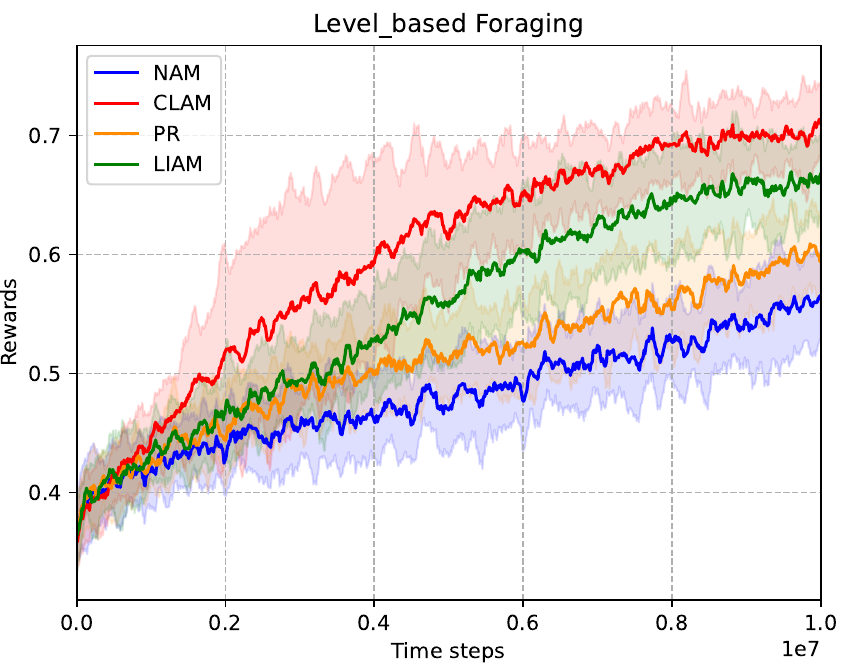}
}
\subfloat{\label{Fig4.sub.2}
\includegraphics[width=0.23\textwidth]{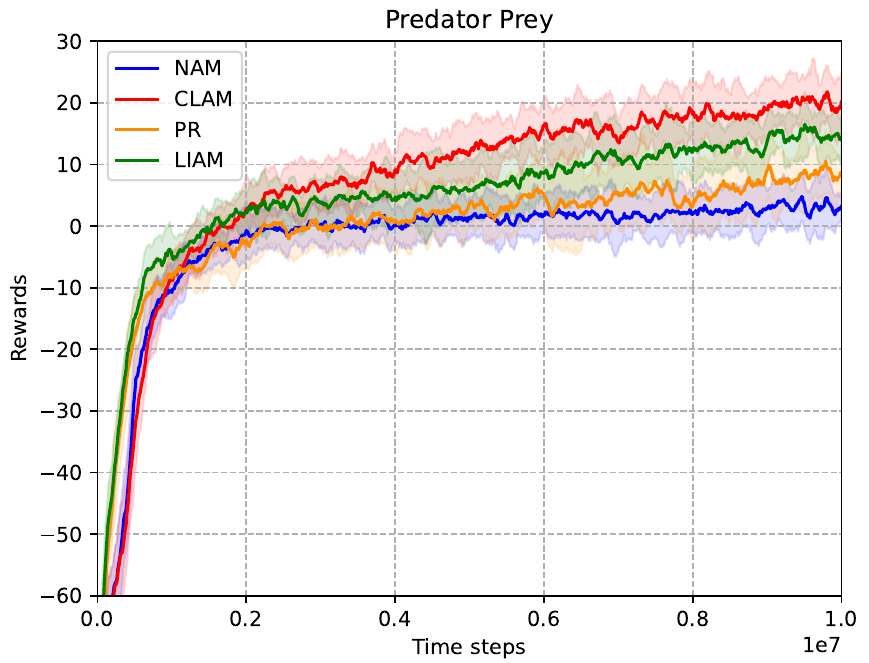}}\\
\subfloat{
\label{Fig4.sub.3}
\includegraphics[width=0.235\textwidth]{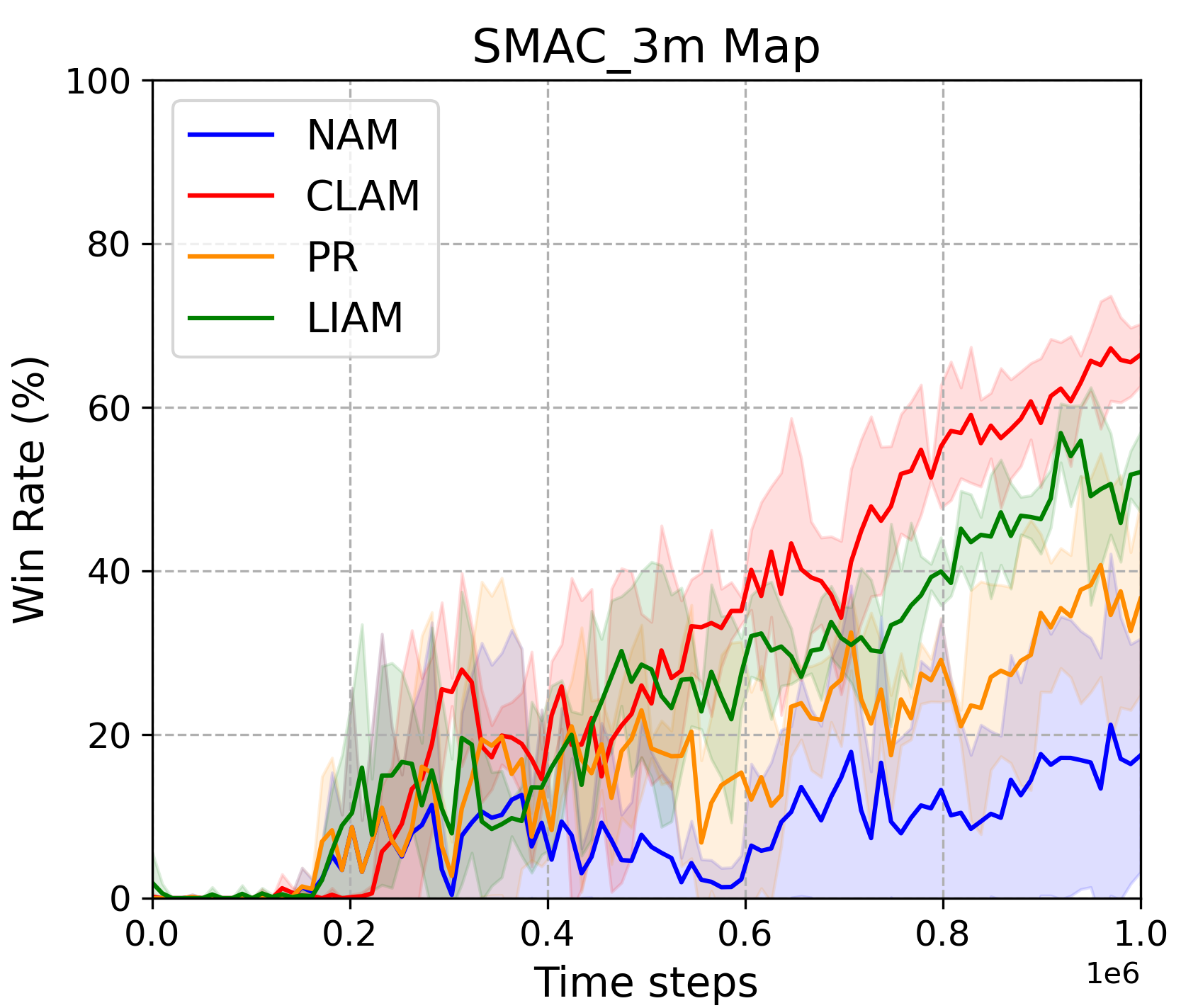}
}
\subfloat{
\label{Fig4.sub.4}
\includegraphics[width=0.235\textwidth]{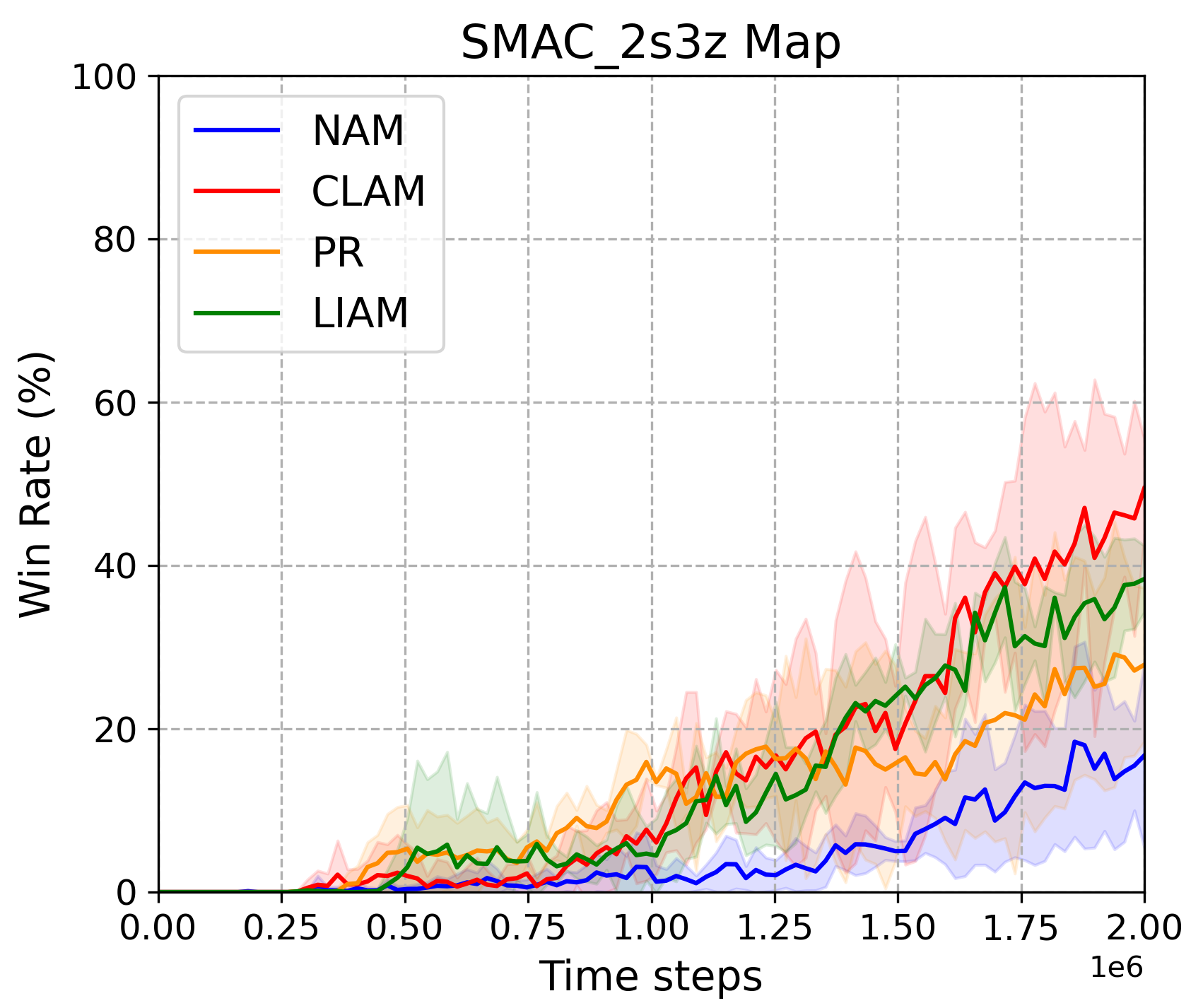}
}
\caption{Episodic evaluation returns and 95 \% confidence interval of the four evaluated methods in four scenarios.}
\label{Fig4}
\end{figure}

\subsection{Baseline Methods}
We chose two widely recognized and representative methods in the field of agent modeling as baselines for our comparative experiments. Additionally, we included a na\"ive PPO method that does not involve agent modeling as another baseline.

\paragraph{No Agent Modeling (NAM)}
Since this approach does not employ any agent modeling techniques, we expect this to be the worst of the methods. NAM merely serves as a benchmark to evaluate whether agent modeling methods do anything to enhance the ego agent's policy performance and, if so, to quantify the degree of that improvement.

\paragraph{Policy Representation Learning in Multi-Agent System (PR)}
This baseline is based on the method proposed by \cite{grover2018learning}. The method is built on an encoder-decoder architecture, where the encoder embeds the input observation trajectory into feature vectors in a point-wise manner, yielding vectors for each time step within the trajectory. The feature vectors are aggregated using the average pooling method to obtain the trajectory's representation. This approach generally requires modeling an ego agent's complete observation trajectory from a previous episode to generate the modeled agent's policy representation.

\paragraph{Local Information Agent Modeling (LIAM)}
This baseline was introduced by \cite{liam}. Like PR, it employs an encoder-decoder architecture. But unlike PR, the encoder is a recurrent encoder. The encoder takes in the history of the ego agent's observations and actions up to the current time step and outputs the modeled agent's policy representation $c_t$ in real time during the current episode. This method requires obtaining the modeled agent's observations and actions for the reconstruction task during the training phase but only needs the ego agent's observations and actions during the execution.

\subsection{Policy Evaluation}
Fig. \ref{Fig4} shows the average return curves for the three baselines and CLAM in four multi-agent environments. We evaluated each method using five different random seeds. The solid lines represent the average return values over five separate runs, while the shaded regions indicate the 95\% confidence interval.
From the figure, it is evident that the NAM baseline performs the worst in all environments, in line with our expectations. This suggests that Na\"ive reinforcement learning algorithms without agent modeling modules cannot acquire any auxiliary information to distinguish between the policies of its opponents or teammates. The PR method performed slightly better than NAM in all environments but fell short compared to the other two methods. We attribute this outcome to the point-wise encoding approach of PR, which does not seem to effectively capture temporal relationships between trajectory steps. This limitation results in so few informative policy representation vectors that the model cannot accurately model the agents' strategy features. Hence, the representation provides limited assistance in the ego agent's adaptive policy training. Additionally, using the average pooling method to aggregate feature vectors might not be the optimal approach.
LIAM, which employs a recurrent encoder, receives the second-highest returns (on average) in all scenarios. This method capitalizes on the temporal information within trajectories for effective representation learning, resulting in policy representations with richer information than PR. As can be seen, the performance improvement is substantial compared to the other two baselines.
CLAM achieved the highest returns in all environments and improved na\"ive reinforcement learning performance by at least 28\%. Especially in more complex environments like SMAC, where the number of agents is larger and the environment settings are more complex, the performance improvement brought by agent modeling methods in policy learning becomes even more significant. In both SMAC maps, the win rate of the CLAM method is approximately three times that of NAM, demonstrating that CLAM effectively and accurately models different adversarial policies. We reason this is due to the attention mechanism, which captures long-range temporal patterns within the trajectory sequences. Moreover, CLAM employs the mask method in the sample augmentation step of contrastive learning. This enhances the model's capability to identify the most distinctive feature information within the trajectories of different policies, resulting in superior performance.

\begin{figure}[!t]
    \centering
    \subfloat[]{\includegraphics[width=0.22\textwidth]{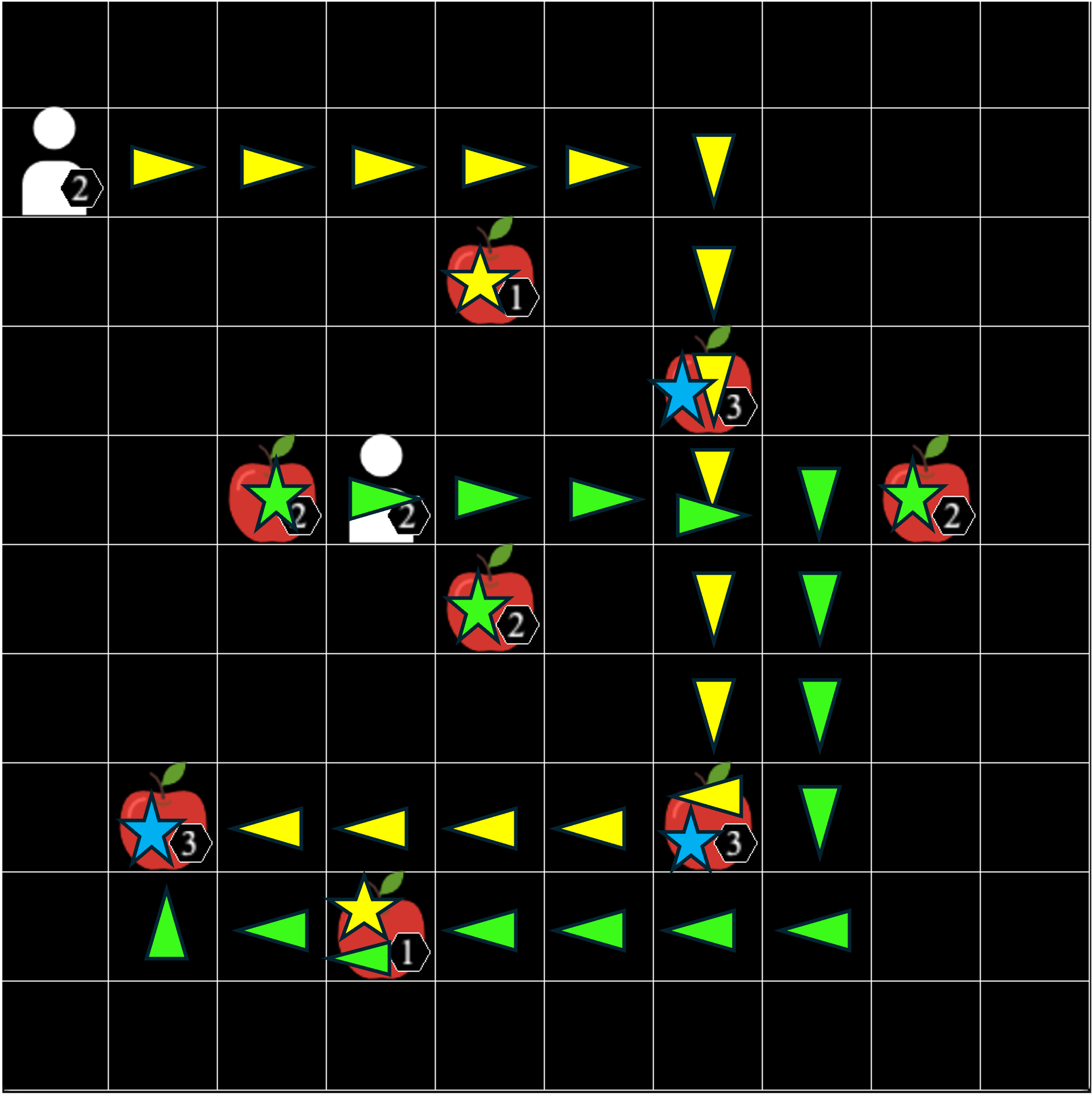}\label{coop}}
    \hspace{0.1cm} 
    \subfloat[]{\includegraphics[width=0.22\textwidth]{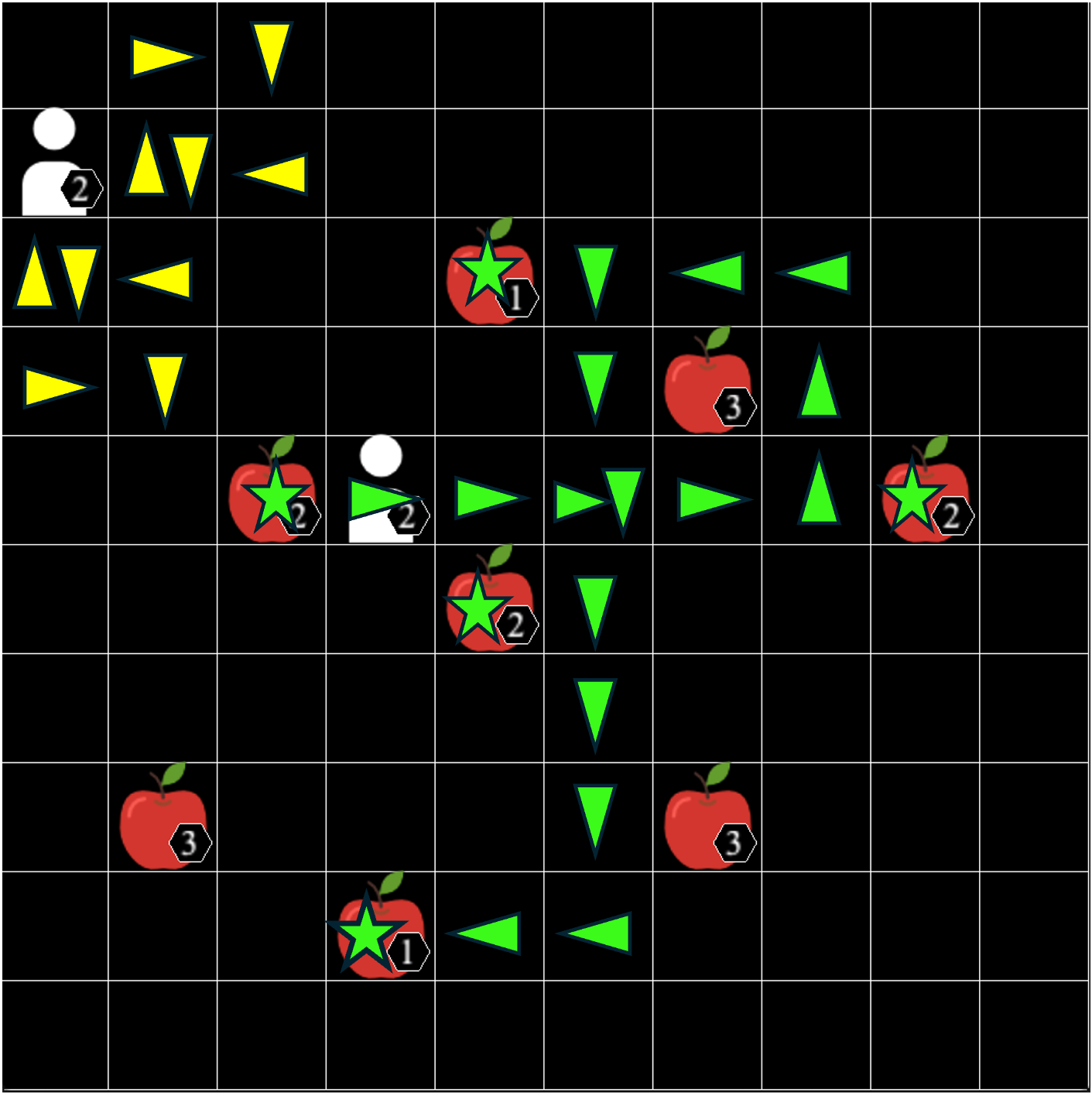}\label{random}}
    \caption{The trajectories of the ego agent collaborating with companions using different policies. (a) The teammate always chooses to collect the closest apple. (b) The teammate executes a random policy. The green arrow represents the action trajectory of the ego agent, while the yellow arrow represents the teammate. The green stars indicate apples collected solely by the ego agent, the yellow stars indicate those collected solely by the teammate, and the blue stars indicate apples collected through cooperation.}
    \label{traj}
\end{figure}

To further illustrate how the ego agent adjusts its policy based on the behavior of teammates employing different policies, we visualized the trajectories in the level-based foraging scenario for two representative policies, as shown in Fig. \ref{traj}. The green arrow trajectory represents the actions of the ego agent, while the yellow arrow trajectory represents the actions of teammates. In Fig. \ref{coop}, the teammate executes a policy of always choosing to collect the closest apple. In this case, the ego agent initially selects apples that are accessible nearby. Subsequently, when it observes the teammate waiting beside apples that require cooperation, the ego agent chooses to collaborate and follows the teammate to jointly collect all the apples that can only be harvested cooperatively. Fig. \ref{random} demonstrates how the ego agent responds to a teammate executing a random policy. From the trajectories, we can see that upon noticing the teammate performing random actions, the ego agent does not choose to follow the teammate or wait beside the apples requiring cooperation. Instead, the ego agent independently collects all apples within its capability to maximize team rewards. This trajectory visualization shows that the ego agent, equipped with the agent policy modeling module, can dynamically adjust its policy in the same scenario based on the policies of different teammates, thereby maximizing team rewards.
\begin{figure}[!t]
      \centering
      \subfloat[]{\includegraphics[width=0.26\textwidth]{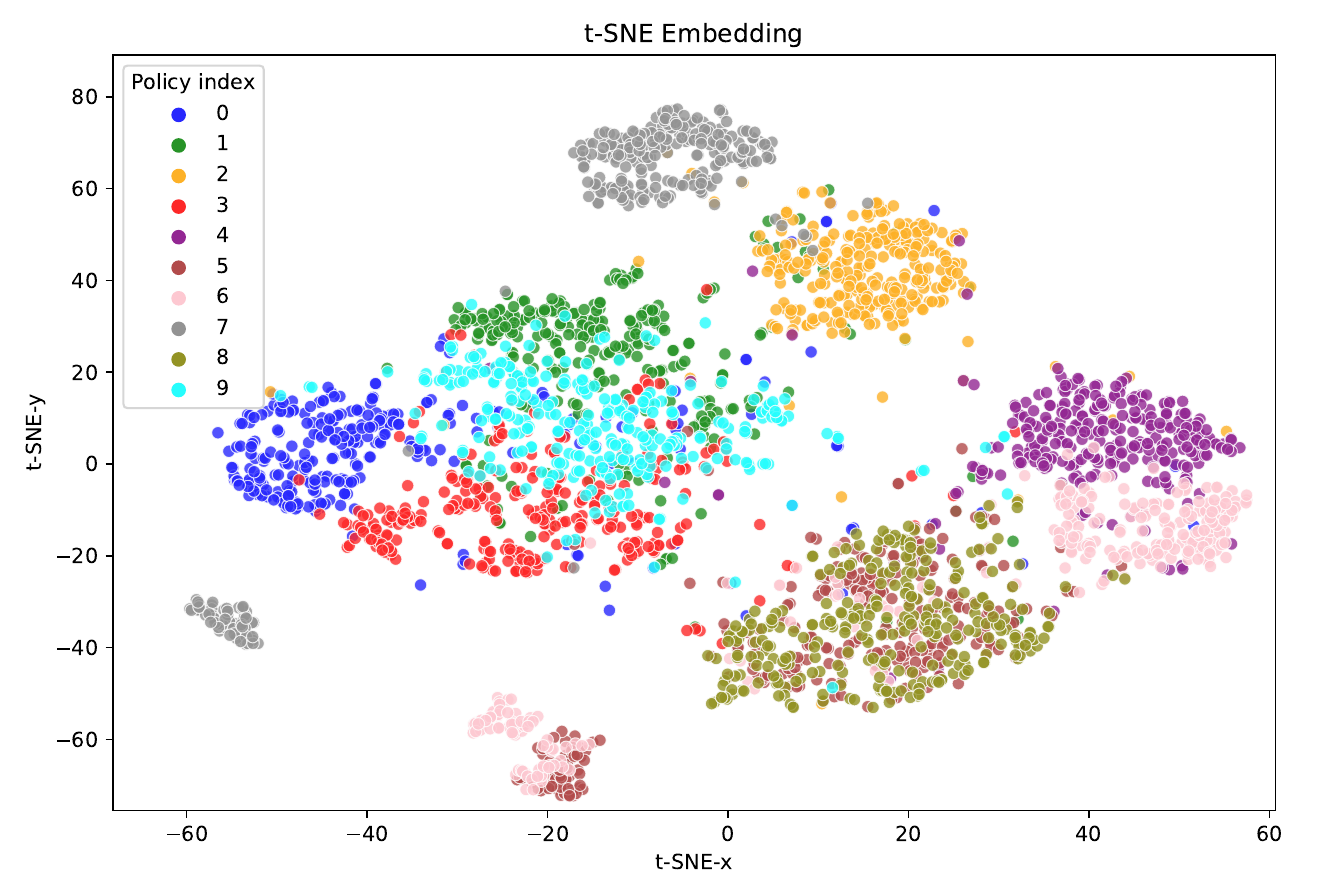}\label{Fig5.sub.1}}
      \subfloat[]{\includegraphics[width=0.24\textwidth]{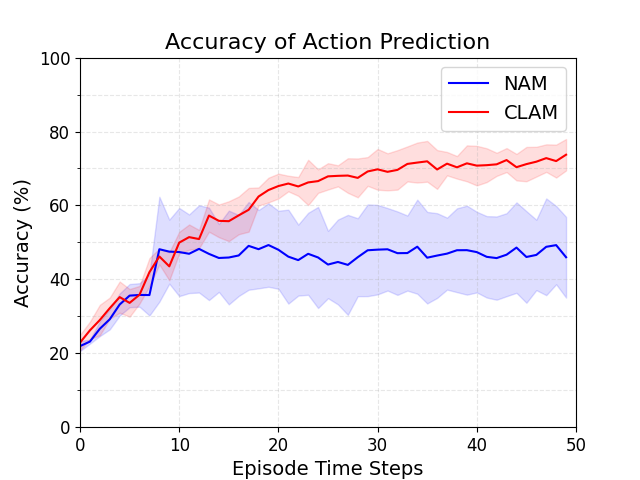}\label{Fig5.sub.2}}
    \caption{(a) t-SNE projection of the embeddings in the predator-prey environment, where different colors represent different fixed policies and different points represent different episodes. (b) The action prediction accuracy. The solid line represents the average accuracy over ten policies, and the shaded region denotes the 95\% confidence interval.}
    \label{Fig5}
\end{figure}

\subsection{Model Evaluation}\label{sec4.d}
Continuing our analysis, we delve into why the CLAM method achieves such remarkable performance. Utilizing the t-SNE method \cite{van2008visualizing}, we projected the policy representation embeddings into a two-dimensional space, as shown in Fig. \ref{Fig5.sub.1}. These were the policy embeddings output by CLAM's encoder at the 25th time step of each episode in the Predator-prey environment. Each color represents a different fixed policy combination for the predator team, and each point represents an episode. From Fig. \ref{Fig5.sub.1}, it is apparent that points of the same color are strongly clustered. This indicates that the CLAM model can identify and differentiate the behavioral characteristics of different predator team policies.

However, we did notice an overlap between Policy 5 and Policy 8. Further examination reveals that these two policy combinations inherently share similar features. As a result, the self-supervised contrastive learning scheme struggled to tell them apart. Nevertheless, this underscores that CLAM can indeed represent policies based on their behavioral similarities, effectively aiding the training of adaptive cooperative or adversarial policies through reinforcement learning.

To further evaluate the effectiveness of the learned policy representations in capturing contextual information, we conduct an agent action prediction experiment in predator-prey environment. As a controlled experiment, we introduce an agent action predictor, composed of MLPs. The predictor is fed with two different input configurations: one combining the real-time policy embeddings $c_t$ generated by the frozen CLAM model with the ego agent observation $o^e_t$, and the other relying solely on the observation $o^e_t$ to predict the actions of modeled agents $a_t^m$ at the corresponding time step $t$. This experiment directly assesses whether the policy embeddings provide sufficient context information to enable the ego agent to gain a deeper understanding of the environment and make more informed decisions.
After training the predictor, we evaluated the action prediction accuracy across observation trajectories of varying lengths from 0 to 49. The results are presented in Fig. \ref{Fig5.sub.2}, where the red line(CLAM) indicates the prediction outcomes when incorporating policy embeddings generated by CLAM, while the blue line (NAM) represents the results based solely on observations.

From the figure, we can see that, at time step 0, both CLAM and NAM's predictive accuracy is near random, approximately 20\%, as the model lacks sufficient context information to infer other agents’ actions. However, as more observations become available, the CLAM's accuracy improves rapidly, reaching around 50\% by step $10$, nearly 65\% at step $20$, and stabilizing above 70\% after step $30$.

In contrast, the baseline NAM shows a slight improvement in accuracy only during the early stages of the episode. However, after approximately $10$ steps, its accuracy stagnates and remains consistently below 50\%. These results demonstrate the CLAM effectively captures contextual information which contains agent-specific behaviors and inter-agent dynamics, allowing for more accurate action predictions. Furthermore, this also proved that policy embeddings learned by CLAM model contribute to reinforcement learning by enabling agents to generalize across different agent policies, leading to more adaptive and robust decision-making in multi-agent environments.

\begin{figure}[!t]
\centering
\includegraphics[width=0.52\textwidth]{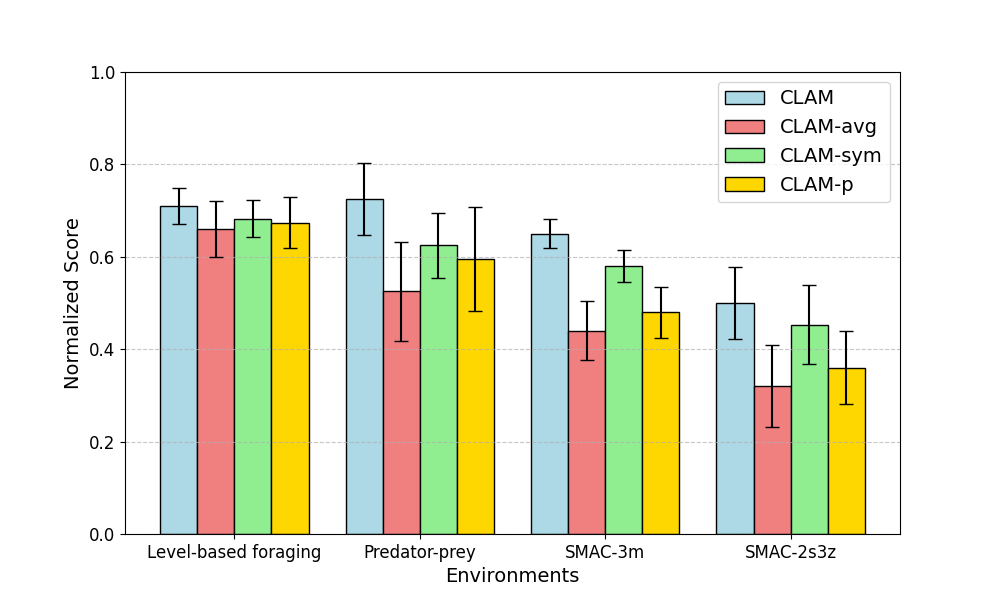}
\caption{Average returns comparison of CLAM against three ablated versions of CLAM in four environments.} \label{fig7}
\end{figure}

\begin{table}
\caption{Intra-Inter Clustering Ratios (IICR) of policy representation embeddings in predator-prey environment.}
\centering
\begin{tabular}{llllll}
\hline
Method $\backslash$ Time step  & 10 & 20 & 30 & 40 & 50  \\
\hline
CLAM     & \textbf{0.89}  & \textbf{0.78}  & \textbf{0.75}  & \textbf{0.74}  & \textbf{0.74} \\
CLAM-\textit{avg}     & 0.94  & 0.88  & 0.85  & 0.83  & 0.82 \\
CLAM-\textit{p}  & 0.91  & 0.87  & 0.83  & 0.82  & 0.80 \\
CLAM-\textit{sym}  & 0.89  & 0.83  & 0.81  & 0.80  & 0.79 \\
\hline
\end{tabular}
\label{tab1}
\end{table}

\subsection{Ablation Study}

This section presents the results of ablation studies conducted to examine the effectiveness of attention pooling and evaluate the impact of asymmetric sample augmentation strategies in our CLAM model. 
\paragraph{Feature aggregation method}
To analyze the effect of different feature aggregation mechanisms, we compare our proposed attention-based pooling with two alternative approaches: (1) CLAM-\textit{avg}, which replaces attention pooling with average pooling, treating all feature vectors equally by computing their mean, and (2) CLAM-\textit{p}, which replaces attention pooling by learning a weight vector $p(1, t)$ to aggregate policy features along the temporal dimension $t$. This alternative design allows us to assess whether a simpler, more computationally efficient learnable vector can achieve a similar effect as attention pooling. Two metrics were leveraged to evaluate the feature aggregation performance: i) the average scores obtained by the agents in the multi-agent environment (the rewards or win rates in all four environments are normalized to the $[0,1]$ range to facilitate comparison and analysis.), and ii) The Intra-Inter Clustering Ratio (IICR) measures the clustering properties of policy embeddings in an embedding space. An IICR value below 1 indicates effective feature representation by grouping similar samples and separating dissimilar ones, with lower values signifying stronger clustering and better representation performance. It serves as a key metric for evaluating the feature learning ability of self-supervised methods and the discriminative power of learned features.

Fig. \ref{fig7} shows that among the three feature aggregation methods, the CLAM method consistently achieves the highest scores across four different scenarios, followed by the CLAM-\textit{p} method, while the CLAM-\textit{avg} method scores the lowest. A similar conclusion can be drawn from Table \ref{tab1}, which presents the IICR results. We evaluated the policy embeddings at time steps $(10, 20, 30, 40, 50)$, and a comparison of the scores clearly shows that the values for CLAM are consistently lower than other methods. We attribute this to the attention pooling module’s ability to better focus on crucial temporal features in the ego agent’s observation trajectories, leading to a more distinct aggregation of feature vectors and ultimately yielding more informative policy representations. The CLAM-\textit{p} method outperforms the CLAM-\textit{avg} method, indicating that the trained weight vector can allocate weights more effectively than simply averaging the feature sequence. However, compared to the attention-based CLAM method, it cannot dynamically adjust weights based on different feature sequence lengths, resulting in lower performance than the attention-based approach.

\begin{figure}[h]
\centering
\includegraphics[width=0.45\textwidth]{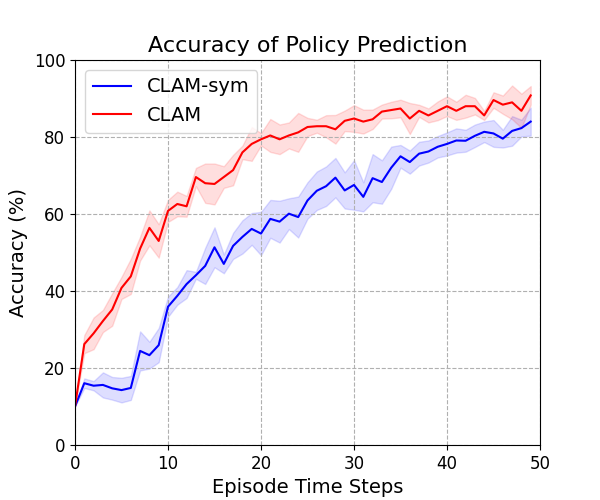}
\caption{Policy prediction accuracy comparison of CLAM against CLAM-sym in predator-prey environment.} \label{fig8}
\end{figure}

\paragraph{Asymmetric sample augmentation method}
Furthermore, to validate the impact of asymmetric sample augmentation approach, we introduced CLAM-\textit{sym}, which removes the masking-based augmentation applied to one side of the contrastive learning sample pairs, resulting in a symmetric sample augmentation setting.
Comparing CLAM and CLAM-\textit{sym} in Fig. \ref{fig7}, CLAM gets higher returns in all four scenarios. The IICR values also suggest that CLAM outperforms CLAM-\textit{sym} in terms of policy representation clustering. To investigate the superiority of our proposed asymmetric sample augmentation method over the symmetric approach, we conducted a policy prediction experiment to further examine the model. Similar to the agent action prediction experiment described in Section \ref{sec4.d}, we froze the trained CLAM encoder and CLAM-\textit{sym} encoder, then attached an MLP-based prediction head to classify policy types based on the learned representation embeddings. We evaluated prediction accuracy across observation trajectories of varying lengths from 0 to 49, with the results shown in Fig. \ref{fig8}. As seen in the figure, there is a significant accuracy gap between CLAM and CLAM-\textit{sym} in the early stage of the episode. CLAM reaches 60\% accuracy within the first 10 time steps, whereas CLAM-\textit{sym} achieves less than 40\%. As the trajectory length increases, the gap between the two methods gradually narrows. We attribute this to the masked-based asymmetric sample augmentation method, which encourages the model to identify similarities among more sparse sample features and capture the most essential and discriminative features. This enhances the model’s prediction accuracy for shorter observation trajectories, thereby improving its ability to quickly adapt to different policies and enhancing the robustness of learned reinforcement learning policies.

\section{Conclusion}
This paper presented a novel agent modeling model that leverages a Transformer encoder and an attention-pooling module. The model effectively captures real-time policy representations of the modeled agents, utilizing only the observation signals from the ego agent. Importantly, our work pioneers the integration of the attention mechanism into the field of agent modeling. This innovative attention mechanism empowers the proposed model to efficiently learn how to capture useful information from the ego agent's local observation, thereby substantially augmenting the model's capacity for representation. Through contrastive learning, and combined with an innovative asymmetric sample augmentation technique for creating positive sample pairs, we successfully trained a policy encoder to model agent policies in a self-supervised manner and encouraged the model to unearth the most distinctive information and features within the ego agent's observation trajectories. Our experimental results demonstrate that CLAM significantly increases the episodic returns obtained by the ego agent compared to baseline methods. More importantly, in contrast to the baseline methods, our approach stands out by eliminating the necessity to acquire observations from the perspective of the modeled agent or to rely on long-range trajectories. It exclusively relies on the observations of the ego agent as the model's input. In future work, we aim to extend the application of the CLAM model to human-machine collaboration environments. In these settings, the CLAM model will process visual inputs by extracting human behavior and environmental features through a vision model. Leveraging this information, the CLAM model will facilitate the training of more intelligent and robust agents. The proposed approach has the potential to enhance the versatility of multi-agent reinforcement learning methods, making them applicable to scenarios involving human-machine collaboration to achieve a common goal, such as in manufacturing or logistics.

\section*{Acknowledgments}
This work was supported in part by the Australian Research Council (ARC) under discovery under Grant DP220100803 and Grant DP250103612, in part by the Australian Cooperative Research Centres Projects under Grant CRCPXI000007, in part by the Australia Defence Innovation Hub under Grant P18-650825, in part by the AFOSR – DST Australian Autonomy Initiative agreement ID10134 and in part by NSW Defence Innovation Network under Grant DINPP2019 S1-03/09.

\bibliographystyle{IEEEtran}
\bibliography{ref}

\vfill

\end{document}